\documentclass{IEEEtran4PSCC}

\usepackage{graphicx}
\graphicspath{ {./figures/} }
\usepackage{booktabs}
\usepackage{makecell}
\usepackage{multirow}
\usepackage{cite}
\usepackage[cmex10]{amsmath}
\usepackage{amssymb}
\usepackage{algorithmic}
\usepackage{multicol}
\usepackage{siunitx}
\usepackage{hyperref}
\usepackage{threeparttable}
\usepackage{url}

\usepackage{xcolor,colortbl}
\definecolor{LightBlueOurs}{rgb}{0.92, 0.95, 0.99}  

\makeatletter
\let\old@ps@headings\ps@headings
\let\old@ps@IEEEtitlepagestyle\ps@IEEEtitlepagestyle
\def\psccfooter#1{%
    \def\ps@headings{%
        \old@ps@headings%
        \def\@oddfoot{\strut\hfill#1\hfill\strut}%
        \def\@evenfoot{\strut\hfill#1\hfill\strut}%
    }%
    \def\ps@IEEEtitlepagestyle{%
        \old@ps@IEEEtitlepagestyle%
        \def\@oddfoot{\strut\hfill#1\hfill\strut}%
        \def\@evenfoot{\strut\hfill#1\hfill\strut}%
    }%
    \ps@headings%
}
\makeatother

\psccfooter{%
        \parbox{\textwidth}{\hrulefill \\ \small{24th Power Systems Computation Conference} \hfill \begin{minipage}{0.2\textwidth}\centering \vspace*{4pt} \includegraphics[scale=0.06]{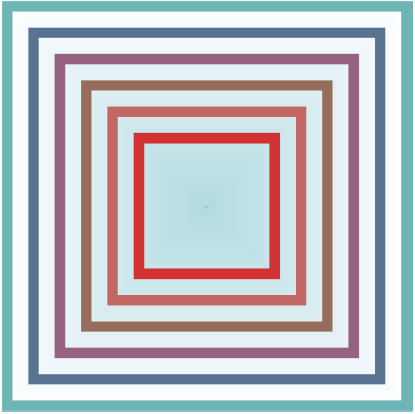}\\\small{PSCC 2026} \end{minipage} \hfill \small{Limassol, Cyprus --- June 8-12, 2026}}%
}

\begin{document}

\title{\huge Selective State-Space Models for Koopman-based Data-driven Distribution System State Estimation}

%% To specify the authors when (number of affiliations <= 2)
\author{\IEEEauthorblockN{Bader Alabdulrazzaq\IEEEauthorrefmark{1}, Bri-Mathias Hodge\IEEEauthorrefmark{1}\IEEEauthorrefmark{2}}
	\IEEEauthorblockA{\IEEEauthorrefmark{1}Department of Electrical, Computer, and Energy Engineering, University of Colorado Boulder,
		Boulder, Colorado 80309}
	\IEEEauthorblockA{\IEEEauthorrefmark{2}Renewable and Sustainable Energy Institute, University of Colorado Boulder,
		Boulder, Colorado 80309}
	\{\tt bader.alabdulrazzaq, brimathias.hodge\}@colorado.edu
}

\maketitle

\begin{abstract}
% 150 words maximum.
Distribution System State Estimation (DSSE) plays an increasingly-important role in modern power grids due to the integration of distributed energy resources (DERs). The inherent characteristics of distribution systems make classical estimation methods struggle, and recent advancements in data-driven learning methods, although promising, exhibit systematic failure in generalization and scalability that limits their applicability. In this work, we propose \textit{MambaDSSE}, a model-free data-driven framework that incorporates Koopman-theoretic probabilistic filtering with a selective state-space model that learn to infer the underlying time-varying behavior of the system from data. We evaluate the model across a variety of test systems and scenarios, and demonstrate that the proposed method outperforms machine learning baselines on scalability, resilience to DER penetration levels, and robustness to data sampling rate irregularities. We further highlight the Mamba-based SSM's ability to capture long range dependencies from data, improving performance on the DSSE task.
\end{abstract}

\begin{IEEEkeywords}
data driven, distribution system, machine learning, state estimation, state-space models
\end{IEEEkeywords}

\begin{NoHyper}
\thanksto{\noindent This work utilized the Alpine high performance computing resource at the University of Colorado Boulder. Alpine is jointly funded by the University of Colorado Boulder, the University of Colorado Anschutz, Colorado State University, and the National Science Foundation (award 2201538).}
\end{NoHyper}

\vspace{-1em}
\section{Introduction}
Distribution System State Estimation (DSSE) plays an increasingly-important role in modern power grids with the integration of distributed energy resources (DERs). DERs transform the distribution system from what was traditionally a unidirectional static system into an active, complex, and dynamic system that requires improved monitoring and control for secure operation~\cite{Dehghanpour_2019}. Distribution systems present unique state estimation challenges due to the limited and sparse measurements that render the system under-determined, topological characteristics that contribute to reduced observability and uniqueness of the system, and generally more complex variable-phase unbalanced systems with larger numbers of nodes, all of which makes classical model-based approaches less effective.

Given these challenges, data driven approaches have emerged as a promising direction of research, further fueled by data availability from the increasing deployment of smart sensor infrastructure such as Advanced Metering Infrastructure (AMI) and Phasor Measurement Units (PMUs). Learning-based methods can be utilized in model-free formulation which eschew the need for physical models of the network~\cite{Markovic_2021}, in hybrid frameworks that integrate both physical and data driven methods~\cite{Hu_2025, DelaVarga_2025}, and as surrogates within optimization frameworks~\cite{Zamzam_2019}. The methods are driven by successes in deep learning-based architectures, many of which have been applied to DSSE in literature, including linear models~\cite{Lave_2019, Azimian_2022}, residual networks~\cite{Bhusal_ResnetD_2021}, graph neural networks (GNN)~\cite{Habib_2024}, physics-aware neural networks~\cite{Zamzam_2020, Ngo_2024}, and sequence-based modeling~\cite{Cao_2020}.

Learning-based methods attempt to address the measurement sparsity in distribution systems by including forecasting approaches~\cite{Coutto_FASE_2009, Huang_2022} either using the previous states directly or by incorporate data from external sources such as historical data or meteorological data, to compensate for the measurement limitations in distribution systems, at the expense of introducing additional uncertainties and errors from the forecasting methods.
Crucially, forecasting using the state history directly may not be suitable for distribution systems due to the highly dynamic system state that is further exacerbated by DERs.
Many methods incorporate a state estimator~\cite{Liu_2020} such as Kalman filters and variants to maintain state estimates across noisy and erroneous inputs, but often require knowledge, at least partially~\cite{Revach_2022}, of the system model.
Models that directly map measurements to system state can perform well on small scale system due to the limited dimentionality and variability in the test case, but often show systematic failure as system sizes and thus complexity increases.
Mestav et al.~\cite{Mestav_2019} incorporates a Bayesian inference approach to the direct mapping models with the addition of a probabilistic filter for bad data detection.
Conversely, models often look to embed additional inductive biases to mitigate these types of bottlenecks.
For example, GNN's have shown favorable results~\cite{Ngo_2024, Yue_2024} as graph-based models naturally provide the appropriate inductive bias to model power systems. They are however sensitive to data quality considerations and rely on network topology knowledge that is a common source of errors in distribution systems~\cite{Geth_2023}.

In this work we present \textit{MambaDSSE}, a model-free state estimation framework that incorporates Selective State Space Models (SSM)~\cite{GuDao_Mamba2023}, a form of Deep SSM models~\cite{Gu_S4_2022}, to capture the underlying temporal behavior of the system, along with a probabilistic state estimator that is linearized end-to-end through a Koopman-theoretic framework. The learning framework combines the modeling capacity of a Mamba-based architecture to learn the complex nonlinear temporal evolution of the system, with a probabilistic estimator that compensates for noisy and bad data to produce a consistent state trajectory.
We evaluate performance across a variety of test cases and scenarios, highlighting the impact of the selective SSM in capturing long-range dependencies to improve performance on the DSSE task, and in adding robustness to data sampling rate perturbation. We evaluate how the choice of learning the underlying system behavior, as opposed to learning measurement to state mapping, allows the model to gracefully scale to large systems.

\begin{figure*}
\centering
\includegraphics[width=1\textwidth, height=6cm, keepaspectratio]{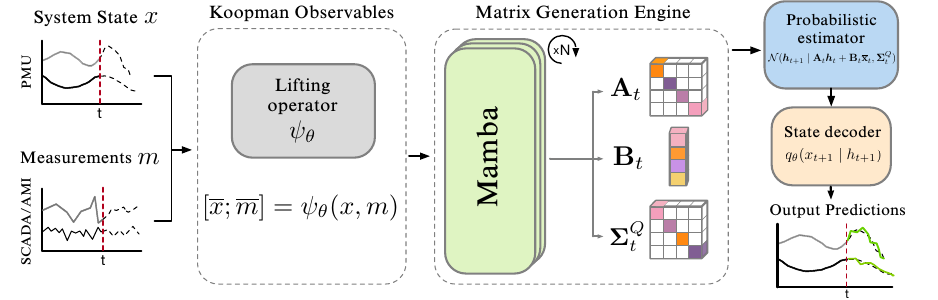} 
\caption{\looseness=-1Overview of the \textit{MambaDSSE} framework. Historical system data is used to train the model to infer the underlying time-varying behavior of the observed system. The input data is lifted into a high-dimensional latent space via learned Koopman observables, to linearize the complex nonlinear dynamics. 
The Mamba-based matrix generation engine produces the state transition matrices that drive a probabilistic state estimation filter that's responsible for modeling the measurement noise and process uncertainty.
A decoding pipeline, conditioned on the posterior state distribution of the filter, produces the final state predictions.}
\label{fig:model_overview}
\end{figure*}

\section{Preliminaries}

\subsection{Selective State Space Models}
Deep SSMs have recently garnered attention as an alternative to the quadratic complexity of Transformer-based models~\cite{Vaswani_2017} for modeling sequential data sources. SSMs, in general, can be described as a continuous-time (CT) linear ODE that map a one-dimensional input $u(t) \in \mathbb{R}^D$ to an output vector $y(t) \in \mathbb{R}^D$ through a latent hidden state $x(t) \in \mathbb{R}^N$, where the CT matrices $\mathbf{A}, \mathbf{B}, \mathbf{C}, \mathbf{D}$ specify the state transition processes and the subsequent output mapping of the process:

\begin{equation}
    \begin{gathered}
        x'(t) = \mathbf{A}(t)x(t) + \mathbf{B}(t)u(t) \\
y(t) = \mathbf{C}(t)x(t) + \mathbf{D}(t)u(t)
    \end{gathered}
\end{equation}

However, deep SSMs have been historically limited by their computational efficiency, in addition to the inherent challenges of training recurrent forms. The emergence of a class of Structured SSMs~\cite{Gu_S4_2022} addresses said limitations through a disciplined model formulation covering parametrization~\cite{Gupta_DSS_2022, Gu_S4D_2022}, initialization~\cite{GuDao_HIPPO_2020}, and discretization~\cite{Gu_TrainHiPPO_2023}. The discretization process transforms continuous-time inputs to discrete representations of the SSM suitable for deep learning paradigms. The discretization converts continuous parameters $A, B$ to discrete counterparts $\overline{A}, \overline{B}$, and while other forms exists, a common discretization form uses the zeroth-order hold (ZOH) as:
\begin{equation}\label{eqn:ssmdelta}
    \overline{\mathbf{A}} = \exp(\Delta A), \qquad \overline{\mathbf{B}} = (\Delta A)^{-1}(\exp(\Delta A) - I) \cdot \Delta B
\end{equation}

where $\Delta$ is a learnable timescale parameter of a linear time-invariant SSM, yielding a discretized form: 
\begin{equation} \label{eqn:discreteSSM}
    \mathbf{x}[k] = \overline{\mathbf{A}}\mathbf{x}[k-1] + \overline{\mathbf{B}}\mathbf{u}[k], \quad \mathbf{y}[k] = \mathbf{C}\mathbf{x}[k]
\end{equation}

Equation \ref{eqn:discreteSSM} can be realized as a parallelizeable global convolution, giving the architecture the benefits of both a linear recurrent (efficient inference) and convolutions (efficient training) contingent on the appropriate structure of the kernel elements~\cite{Gu_LSSL_2021}. The convolutional form is expressed as:
\begin{align*}
    \boldsymbol{\overline{K}} = (\overline{CB}, &\overline{CAB}, \dots, \overline{CA^{L-1}B}) \\
    &y = \boldsymbol{\overline{K}} * u
\end{align*}

In Mamba~\cite{GuDao_Mamba2023}, the authors improve the expressiveness of the Structured SSM by eliminating the time-invariant formulation and introducing a data-dependent parametrization of $\mathbf{B}$, $\mathbf{C}$ and time variable $\Delta$ that significantly improve its performance. The now context-aware time-varying structured matrices enables them to propagate only information deemed relevant at each time step from the input, hence the "selective" SSM connotation of the Mamba model. Mamba-based models have been shown to perform on a wide range of benchmarks and tasks~\cite{Liu_VMamba_2024, Wang_SMamba_2024}, with particularly strong performance in tasks that require capturing long-range dependencies~\cite{Goel_Sashimi_2022, Lu_ICRL_2023}.
In this work, we utilize selective SSMs as a backbone that empowers our dynamic matrix generation module, and empirically show its role in improving the model performance for the DSSE task, leveraging its time-dependent parametrization and capacity to model long range dependencies.

\subsection{Koopman Theory}

Koopman Theory~\cite{Koopman1931} represents nonlinear dynamic systems as a linear system operating on a lifted space of measurement functions $\psi$, such that the complex evolution of the state can be operated on, through a Koopman operator $\mathcal{K}$, as a linear operator on observables. As it is challenging to model complex dynamics directly from nonlinear and noisy states, Koopman operators map the nonlinear system state into an infinite-dimensional linear system:
\begin{equation*}
    \psi(x_{k+1}) = \psi \circ f(x_k, u_k) = \mathcal{K}\circ \psi(x_k, u_k) 
\end{equation*}

Eigenanalysis and spectral analysis in the Koopman framework enables us to understand the behavior of the observed nonlinear system. Dynamic Mode Decomposition (DMD) is often used to identify approximate finite-dimensional Koopman operator representations based on linear measurements, where the Extended DMD~\cite{Schmid_2010} lifts the linear assumptions and allows the model to process nonlinear measurements.
Their performance, however, is bottlenecked by the manual selection of a set of basis functions. Finding such functions that can appropriately represent the complex dynamics remains an open problem in itself~\cite{Brunton_2022}.
Data-driven representation learning methods instead utilize end-to-end learning of the lifting transformations to discover nonlinear embeddings to linearize complex systems, strictly from data without knowledge of the system governing equations.
Recent data-driven methods make use of Koopman theory, leveraging measurement function learning, to decompose nonlinear dynamics into linear systems, and to learn invariant subspaces to solve downstream tasks of interest~\cite{Han_2020, liu2023koopa}.

In this work, we apply deep learning methods to learn a high-dimensional approximation of the Koopman observables via learning a lifting operator of invariant subspaces, with the goal of applying linear state estimation for the underlying time-variant nonlinear system.

\subsection{Formulation of Distribution System State Estimation}

The objective of distribution system state estimation is to estimate state quantities of all buses in a given system, given partial and noisy measurements of the network. The system states and measurements are related via a nonlinear measurement function: 
\begin{equation*}
    \mathbf{z} = h(\mathbf{x}) + \boldsymbol{\varepsilon}
\end{equation*}

where the system states $\mathbf{x}\in \mathbb{C}^n$ are mapped into a set of real-time measurements $\mathbf{z}\in \mathbb{C}^m$ via a nonlinear measurement function $\mathbf{h}(\cdot)$---typically nonconvex power flow equations---and measurement noise $\boldsymbol{\varepsilon}$. The system state variables we wish to infer are the unbalanced variable-phase distribution system states represented as voltage phasors $V \in \mathbb{C}^{3\phi N}$.

Distribution systems generally do not have sufficient real-time data to make the system fully observable, in addition to its sparsity and topological characteristics that make traditional power system state estimation methods fail. Measurements typically consist of voltage and current phasors obtained from PMUs, active and reactive power flows, bus injections, voltage and current magnitudes obtained from SCADA systems. Load and generation measurements obtained from smart meters can also be incorporated into the state estimation formulation, with a caveat that these AMI data sources provide delayed measurements which then require additional considerations for the observer to fuse them with the real-time measurements. Historical profiles of the network, load and generation forecasts, and other supplemental data sources can be integrated as "pseudo-measurements" to improve the observability of the system at the expense of a higher noise variance and increased uncertainty in general. 
In this work, we address the limited observability challenges by leveraging the capacity to capture long-range dependencies to enable the model to learn the temporal behavior and underlying correlations of the system under study, given a collection of observations from previous time steps. We show that exploiting such structure from data allows us to overcome the under-determined nature of distribution systems without the need for external data sources, whether in the form of pseudo-measurements or from prior knowledge of the system's topology or governing models.

\section{The \textit{MambaDSSE} Model}
MambaDSSE is a distribution system state estimation framework consisting of a selective SSM model that learns to infer the time-varying behavior of the distribution system from measurement input sequences, and a probabilistic state estimator, linearized by a Koopman operator, that propagates process noise and uncertainty to predict the system states.
The model takes as input a small subset of system states and measurements in a manner that renders the system unobservable, and predicts the remaining states while requiring no knowledge of system models nor the topology of the distribution network.
An overview of the proposed MambaDSSE model and its main components are shown in Fig.~\ref{fig:model_overview}.
The Mamba-based backbone, responsible for learning the systems temporal behavior, functions as a \textit{matrix generation engine} that utilizes an input-dependent SSM capable of capturing long-range dependencies to learn the time-varying nonlinearities of the observed system, described in Section~\ref{sec:SSM_as_mge}. The matrix generation engine drives a probabilistic estimator responsible for learning to model the estimation process uncertainty to produce state predictions, outlined in Section~\ref{sec:PSE}. 
We impose several constraints on the structure of the produces matrices in an effort to maintain the stability of the estimation process and to improve the computational efficiency, which we outline in Section~\ref{sec:ssm_structure}.

\subsection{Selective SSM as a Matrix Generation Engine}
\label{sec:SSM_as_mge}
\looseness=-1The Mamba-based backbone model acts as a matrix generation engine that learns the time-varying behavior of the observed system to produce the state matrices for the linearized state estimator. Following the Mamba architecture, we assign $D$ separate SSM's, one for each individual dimension of the input vector $x \in \mathbb{R}^D$---an inductive bias that resembles the diagonal noise weight matrix in Weight Least Square (WLS) methods traditionally used in power system state estimation.

The generation of the state transition matrix $\mathbf{A}_t = \mathcal{A}(\mathbf{\hat{x}}_{<T},\mathbf{\hat{m}}_{\leq T})$, input control matrix $\mathbf{B}_t = \mathcal{B}(\mathbf{\hat{x}}_{<T},\mathbf{\hat{m}}_{\leq T})$, and the process covariance $\mathbf{\Sigma}^Q = \mathrm{\Sigma^Q}(\mathbf{\hat{x}}_{<T},\mathbf{\hat{m}}_{\leq T})$ are parametrized on the previous observations of the system state $\mathbf{\hat{x}}_{<T}$ and network measurements $\mathbf{\hat{m}}_{\leq T}$. The model inputs are projected into lifted measurement space from a learned measurement function, which in practice we implement as an MLP-based encoder network $[\overline{x}; \overline{m}] = \psi_{\theta} (x, m)$ that is trained end-to-end with the model.

\subsection{Matrix Structure Constraints}\label{sec:ssm_structure}
\looseness=-1We enforce specific structures on the matrices generated by the selective SSM to maintain stability of the system and to enable an efficient computational pipeline. To maintain the stability of the system across training, we enforce the matrix to the Routh-Hurwitz criterion for stability by constraining the state transition matrix, the main component that represents the temporal progression of the system, to a matrix with all negative eigenvalues. The process noise covariance $\mathbf{\Sigma}^Q$ is constrained to be positive semi-definite to maintain a physically-viable covariance matrix. Following prior work on matrix parametrization~\cite{Gu_S4D_2022}, all generated matrices are constrained to diagonal forms which improves the stability and computational efficiency of the model.

\subsection{Learning valid approximations of Koopman observables}
The learned approximations of measurement mappings project input space states into a lifted space constrained to span a Koopman invariant subspace. The model is empowered to learn valid approximations of Koopman observables~\cite{Lusch_2018} by enforcing the learned observables to evolve in a strictly linear system; and ensuring the trained model sufficiently predict and reconstruct future states of the nonlinear dynamics via a linear operator on the learned observables.
The model is optimized end-to-end using a simple mean squared error loss in the form $\Vert \mathbf{x^*}_{k+1} - \psi^{-1}({K}_\theta \circ\psi({\mathbf{x}_k}))\Vert$, where $\mathbf{x^*}$ are the target states and $\mathbf{x}$ are the input states---serving as a prediction and reconstruction loss operating in input space.

\subsection{Probabilistic State Estimator}\label{sec:PSE}
The outputs of the matrix generation engine drive the step process of the linear Gaussian SSM of a Kalman filter state estimation where our model $p_{\theta} \left(\boldsymbol{h}_{t+1} \mid \boldsymbol{h}_t, \mathbf{\overline{x}}_t \right)$ is a parametrization of a Gaussian distribution $\mathcal{N}(\boldsymbol{h}_{t+1} \mid \mathbf{A}_t\boldsymbol{h}_t + \mathbf{B}_t\mathbf{\overline{x}}_t, \boldsymbol{\Sigma}^Q)$. We also learn a time-invariant diagonal measurement noise matrix $\mathbf{R}$ from the data without any prior knowledge of the state process. The aforementioned constraints on the generated matrix structure improves the computational efficiency of the state estimation, as the diagonal structure transforms the state update steps into element-wise operations and allows us to substitute the matrix inversion in the update step with a scalar division operation. 
Finally, a decoding pipeline, parametrized by the mean and covariance of the probabilistic estimator, is used to produce input-space system states predictions.

\section{Experimental setup}

\subsection{Experiments \& Ablations Studies}

\textbf{Learning the underlying temporal behavior to alleviate scalability bottlenecks}
\;\;Models that directly map measurements to states struggle as systems scale and complexities increase. We investigate whether empowering the model to instead learn the time-varying behavior of the system leads to performance improvements on larger scale systems.
The models are evaluated on test cases ranging from a 71-bus to a 6746-bus distribution test system, with a fixed sequence length of 72 hours. Detailed descriptions of the test systems used in the analysis are presented in Section~\ref{sec:case-studies}.

\textbf{Sensitivity towards DER penetration levels}
While we emphasize test scenarios with high generation variability in our analysis in general, state estimation in practice requires resilience towards distribution-side generation profiles, given the diverse operational characteristics brought on by the different DER percentages. To examine this sensitivity, we test models on multiple generation scenarios and examine the performance threshold on each state parameter as we vary the DER penetration levels.

\textbf{Role of long-range dependency modeling in DSSE}
We emphasize incorporating the ability to capture long range dependencies in the data as an inductive bias that drives our matrix generation model.
This hypothesis is explored by examining the impact of varying the data sequence length for the model, where we aim to highlight two distinct insights. First, the role that the ability to capture long range dependencies plays in the DSSE task. Second, the importance of incorporating such capacity into the model by design, where in our case we show how the Mamba-based backbone plays a vital role on the downstream task.

\textbf{Importance of probabilistic estimator}
We investigate the benefits of incorporating a probabilistic state estimator for DSSE. In this ablation study, we introduce "duals" of models where one version integrates the state estimator while the other version incorporates a decoder that maps the latent space directly into the system state space. Further discussion on the models and baseline under study in this work are presented in Section~\ref{sec:baselines-models}.

\subsection{Case studies}
\label{sec:case-studies}

\begin{table}
  \centering
  \caption{Overview of SMART-DS test systems}
  \label{tab:smartds_testcases}
  \begin{tabular*}{\columnwidth}{@{\extracolsep{\fill}} lrrr}
    \toprule
    \textbf{Feeder name} & {$\boldsymbol{\vert \mathcal{N}\vert}$} & \textbf{No. PV }& \textbf{No. customers}\\
    \midrule
    % p35uhs0\_4-p35udt1       &      &      &      &      \\
    \texttt{p4uhs0\_4-p4udt4}         & 71   &  17    &   20   \\
    \texttt{p5rhs0\_1247-p5rdt191}    & 435  &  125   &  148   \\
    \texttt{p1uhs0\_1247-p1udt1470}   & 1573 &  632   &  743   \\
    \texttt{p24uhs3\_1247-p24udt6054} & 6746 &  3090  & 3633   \\
    \bottomrule
  \end{tabular*}
\end{table}

\looseness=-1The test systems used in this work are obtained from the publicly-available SMART-DS~\cite{SMARTDS_Dataset_2981} dataset, from which we select a diverse set of subsystems and feeders to test the state estimation performance of the models and outlined experiments. The data is generated from quasi-static time series simulations using OpenDSS, and spans a three year period starting from 2016 using 15-minute resolution, resulting in 35k data points per year. An overview of the selected test systems is shown in Table~\ref{tab:smartds_testcases}.
Scenarios with the largest share of DERs are selected for analysis, unless stated otherwise and DER percentages are referenced by the assigned SMART-DS references (base, low, medium, high, extreme). For the experiments pertaining to the scalability of the models, we fix the input data length to 24 hours and vary the system size. For the DER penetration experiments and the long-range dependency experiments we use the 435-bus system, where the PV penetration levels are varied for the former and the input length is varied for the latter.

\subsection{Data Preprocessing}
The simulation data is split into training and test set using two years of data for training and one year for the test set. The models input vectors are constructed by concatenation of the voltage and angle state measurements along with the sampled active and reactive power injections and line currents. Measurement noise is added to the input data modeled as zero-mean Gaussian noise with standard deviation for each measurement type. PMU voltage and current phasor measurement noise are randomly selected between 1\% to 3\% for each bus individually, and active and reactive power measurements are assigned noise with 5\% standard deviation.
Throughout the analysis in the paper, we assume low observability conditions where the visible buses in the system are insufficient to make the system fully observable. The observable buses are selected by randomly sampling 10\% of the system buses, rather than selecting best-case measurement placement.
For analysis that involves higher data resolution than the default 15 minutes, we modify the load and generation profiles for the given scenario and generate new time-series simulations, while freezing other system configurations.

\subsection{Baseline Models}\label{sec:baselines-models}
We compare the performance of our model against two main model categories: sequence models and dense mixer models, that are purely data-driven and exclude any priors of the network. 
We aim to cover a comprehensive selection of data-driven learning architectures, suitable for the state estimation task at hand. 
We use a Mamba-based baseline, which we refer to as \textit{Mamba Mixer}, in a seq-to-point mixer configuration which substitutes the state estimator we use in our MambaDSSE with a decoder that directly transforms latent space representations into system states outputs. 
Similarly, we use an LSTM~\cite{Hochreiter_LSTM_1997}, first as a residual~\cite{He_ResNets_2016} recurrent neural network baseline (referred to as \textit{LSTM} henceforth), and second, as an in-place variation of the selective SSM. This baseline, which we refer to as \textit{LSTM+SE}, is the same configuration as our MambaDSSE model, except with the Mamba backbone replaced with an LSTM. 

\subsection{Training \& Evaluation}

The training pipeline consists of a supervised setup using a mean squared error loss optimized Adam~\cite{Kingma_Adam_2017} with a learning rate of $1e^{-3}$, weight decay of $1e^{-2}$, and a batch size of 64. We also schedule the learning rate to stabilize the training with linear scaling warm-up for 8\% of the steps and a cosine decay schedule after the initial warm-up period. We use a latent dimension of 256/512 for the Mamba/LSTM sequence processing modules, and the dimension of the lifted Koopman space is determined by an expansion factor on the input space dimension ranging from 1.5-2x depending on test system size. All models are scaled to equalize the number of parameters, while maintaining the dimensionality of the Koopman measurement functions across models.

The performance of each model is evaluated on two common regression metrics, the root mean squared error (RMSE), and the mean absolute error (MAE):
\vspace{-0.5em}
\begin{align}
    &\text{RMSE} = \frac{1}{N} \sum_{i=0}^N (y_i - \hat{y}_i)^2, \quad \text{MAE} = \frac{1}{N}\sum_{i=0}^N \vert y_i -\hat{y}_i\vert
\end{align}

Additionally, we are interested in the errors associated with the different types of predicted parameters in the network state. The same metrics are also used to calculate phasor magnitude errors (in p.u.) and angles (in radians) individually.

\section{Results and discussion}
\begin{table*}[t] 
\caption{Comparison of MambaDSSE and baselines performance on various system sizes$^*$}
\begin{tabular}{lrrrrrrrr}
\toprule
 & \multicolumn{2}{c}{\textbf{71 bus}$^{\dag}$} & \multicolumn{2}{c}{\textbf{435 bus}$^{\dag}$} & \multicolumn{2}{c}{\textbf{1573 bus}$^{\ddag}$} & \multicolumn{2}{c}{\textbf{6746 bus}$^{\ddag}$} \\
\cmidrule(lr){2-3} \cmidrule(lr){4-5} \cmidrule(lr){6-7} \cmidrule(lr){8-9}
& \makecell{MAE \\ {\scriptsize ($\times10^{-3}$)}} & \makecell{RMSE \\ {\scriptsize ($\times10^{-2}$)}} & \makecell{MAE \\ {\scriptsize ($\times10^{-3}$)}} & \makecell{RMSE \\ {\scriptsize ($\times10^{-2}$)}} & \makecell{MAE \\ {\scriptsize ($\times10^{-3}$)}} & \makecell{RMSE \\ {\scriptsize ($\times10^{-2}$)}} & \makecell{MAE \\ {\scriptsize ($\times10^{-3}$)}} & \makecell{RMSE \\ {\scriptsize ($\times10^{-2}$)}} \\
\midrule
LSTM &  5.797 \scriptsize{$\pm$ 0.488}& 6.495 \scriptsize{$\pm$ 0.253}  & 6.137 \scriptsize{$\pm$ 0.205}  & 6.793 \scriptsize{$\pm$ 0.090} & 6.642 \scriptsize{$\pm$ 0.019} & 6.684 \scriptsize{$\pm$ 0.020} & 9.789 \scriptsize{$\pm$ 0.032} & 10.927 \scriptsize{$\pm$ 0.073} \\
LSTM+SE & 5.554 \scriptsize{$\pm$ 0.301}  & 6.130 \scriptsize{$\pm$ 0.260} & 5.686 \scriptsize{$\pm$ 0.375} & 6.553 \scriptsize{$\pm$ 0.139} & 6.249 \scriptsize{$\pm$ 0.381} & 6.761 \scriptsize{$\pm$ 0.108} & 8.921 \scriptsize{$\pm$ 0.005} & 9.911 \scriptsize{$\pm$ 0.035} \\
Mamba Mixer & 4.322 \scriptsize{$\pm$ 0.039} & 6.083 \scriptsize{$\pm$ 0.231} & 5.019 \scriptsize{$\pm$ 0.126}  & 6.220 \scriptsize{$\pm$ 0.122} & 5.439 \scriptsize{$\pm$ 0.336} & 6.277 \scriptsize{$\pm$ 0.041} & 7.677 \scriptsize{$\pm$ 0.216} & 9.219 \scriptsize{$\pm$ 0.084} \\
\rowcolor{LightBlueOurs} MambaDSSE {\scriptsize (Ours)} & \textbf{3.174 \scriptsize{$\pm$ 0.225}}& \textbf{5.949 \scriptsize{$\pm$ 0.037}} &\textbf{3.283 \scriptsize{$\pm$ 0.201}}& \textbf{6.117 \scriptsize{$\pm$ 0.105}} &\textbf{3.804 \scriptsize{$\pm$ 0.125}} & \textbf{6.154 \scriptsize{$\pm$ 0.035}} & \textbf{5.634 \scriptsize{$\pm$ 0.187}} & \textbf{9.039 \scriptsize{$\pm$ 0.046 }}\\
\bottomrule
\end{tabular}
\begin{tablenotes}
        \item[*] $^*$ values reported as mean $\pm$ std.\;\;\;\;${\dag}$ denotes experiments performed on five seeds. \;\;\;${\ddag}$ denotes experiments performed on three seeds.
    \end{tablenotes}
\label{tab:bus_size_model_comparison}
\end{table*}

We analyze the performance of our proposed model on test cases at various scales to highlight the capacity of models that learn underlying system behavior rather than direct mappings that are prone to extraneous representations that suffer loss of performance in large-scale systems. The results are shown in Table~\ref{tab:bus_size_model_comparison}. We note that the proposed model outperforms all other methods compared, across all test system scales, highlighting the improved model capacity introduced by the learned behavior of the system. MambaDSSE demonstrates a 26.5\% reduction in MAE compared to next best performer on the smaller systems and a 30\% reduction on larger systems. Second, we note that the models that utilize a state estimator perform better than their counterparts that use direct latent-to-state mappings, and models that use a Mamba-based architecture also outperform the other backbone architectures, highlighting the benefits of the architectural choice of MambaDSSE that combines both the selective SSM and a probabilistic state estimator.

\begin{figure}
\centering
\includegraphics[width=1.0\columnwidth]{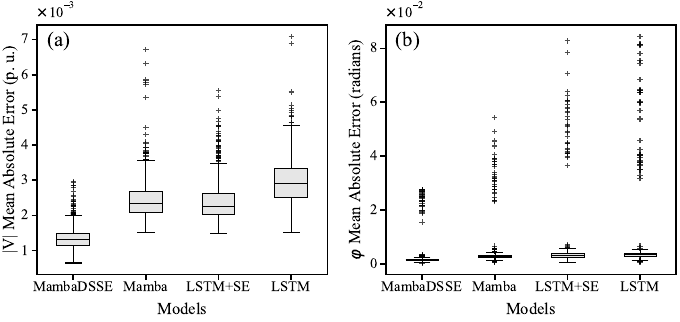}
\caption{Distribution of state prediction errors across all buses of a 435-bus system, for (a) the voltage magnitude, (b) the voltage angle}
\label{fig:boxplot_errors}
\end{figure}

The MambaDSSE RMSE value tends to remain consistent across system sizes, while also being higher than MAEs in general, which is an indicator that the model errors are not uniform across state variables. The distribution of prediction errors across all buses isolated by system state parameter type, voltage magnitude and angle, are shown in Fig.~\ref{fig:boxplot_errors}. Angle errors tends to have larger magnitude and are prone to outliers that can be orders of magnitude above the median, which are then reflected in the the RMSE metric giving more weight to larger errors. Voltage magnitude errors tend to have a tighter concentration. Under both parameters our proposed model outperforms the baseline models in terms of quantitative metrics as well as comparative improvement in performance on the buses with larger errors. To address outliers and persistent erroneous parameters, future work may improve on the loss function by assigning more weight to these specific parameters.

\begin{figure}
    \centering
    \includegraphics{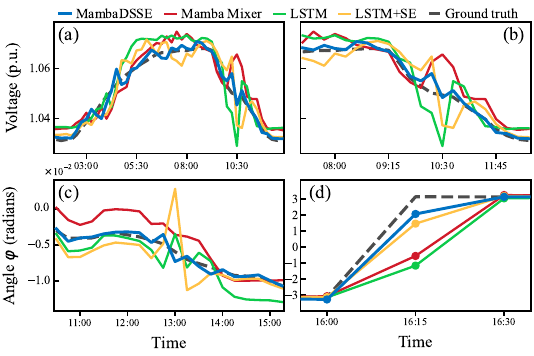}
    \caption{Comparison of model predictions and tracking during system state transitions. (a)-(b) voltage magnitude transitions, and (c)-(d) phase angle transition}
    \label{fig:Vang_performance_sample_onecol}
\end{figure}

Fig.~\ref{fig:Vang_performance_sample_onecol} illustrates the model predictions across a sample of buses, highlighting periods of large errors. The sub-figures (a) and (b) showcase variations in the voltage magnitude during a window of a voltage rise and the transition into and away from stable voltage values. In both instances we note the superior performance of MambaDSSE, especially at detecting changes in trajectory which are the times where the majority of high errors occur.

In Fig.~\ref{fig:Vang_performance_sample_onecol}.(c) we note the prediction stability that the selective SSM backbone affords the models, where both the Mamba-based models showcase smooth and consistent predictions with our proposed model providing predictions that are additionally closer to the ground truth. Fig.~\ref{fig:Vang_performance_sample_onecol}.(d) shows a segment of angle shift with higher angular separation from the reference angle, where the propagation and update step of estimated state in the probabilistic state estimator helps improve the models predictions at these transition periods.

\begin{table*}[htbp]
  \centering
  \caption{Comparison of model performance across different DER scenarios (MAE $\times 10^{-3}$)}
  \label{tab:DER_models}
  \setlength{\tabcolsep}{8pt}
  \begin{tabular}{l *{10}{S[table-format=1.3]}}
    \toprule
    \textbf{Scenario} & \multicolumn{2}{c}{\textbf{Base {\scriptsize(0\% DER)}}} & \multicolumn{2}{c}{\textbf{Low}} & \multicolumn{2}{c}{\textbf{Medium}} & \multicolumn{2}{c}{\textbf{High}} & \multicolumn{2}{c}{\textbf{Extreme}} \\
    \cmidrule(lr){1-1} \cmidrule(lr){2-3} \cmidrule(lr){4-5} \cmidrule(lr){6-7} \cmidrule(lr){8-9} \cmidrule(lr){10-11}
    \textbf{Model} & {$\vert \text{V}\vert$} & {$\varphi_v$} & {$\vert V\vert$} & {$\varphi_v$} & {$\vert \text{V}\vert$} & {$\varphi_v$} & {$\vert \text{V}\vert$} & {$\varphi_v$} & {$\vert \text{V}\vert$} & {$\varphi_v$} \\
    \midrule
    LSTM  & 0.973 & 0.781 & 0.995 & 0.809 & 1.075 & 0.728 & 2.396 & 9.359 & 2.983 & 10.617 \\
    LSTM+SE & 0.393 & 0.314 & 0.626 & 0.490 & 0.767 & 0.724 & 2.256 & 8.842 & 2.603 & 9.312 \\
    Mamba Mixer & 0.453 & 0.351 & 0.511 & 0.374 & 0.760 & 0.606 & 2.029 & 6.704 & 2.251 & 8.395 \\
    \rowcolor{LightBlueOurs} MambaDSSE {\scriptsize (Ours)} & {\bfseries 0.263} & {\bfseries 0.211} & {\bfseries 0.325} & {\bfseries 0.231} & {\bfseries 0.421} & {\bfseries 0.337} & {\bfseries 1.306} & {\bfseries 4.725} & {\bfseries 1.351} & {\bfseries 4.934} \\
    \bottomrule
  \end{tabular}
\end{table*}
\begin{figure}
    \centering
    \includegraphics[width=0.95\linewidth]{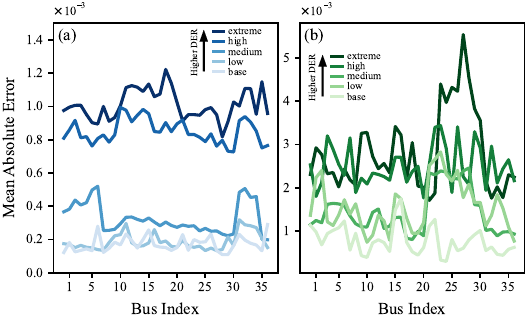}
    \caption{Prediction error (MAE) across different DER penetration scenarios for (a) the proposed model, (b) LSTM+SE baseline}
    \label{fig:DER_comparison}
\end{figure}

We conduct a sensitivity test on the impact of different DER penetration levels, starting from the base case of 0\% DER for the system. The results are shown in Table~\ref{tab:DER_models}. Our proposed method outperforms all baselines in both voltage magnitude and angle errors, and remains consistent as we vary the DER levels.
The relative ease of the task at low DER levels yields low errors in general, which then increase as DERs, and thus task complexity, increases.
At the highest studied DER levels, MambaDSSE error is 40\% lower than the second best model. Equally importantly, the model design choices emerge not only in the raw performance metrics, but also in the the qualitative consistency of the predicted values. Fig.~\ref{fig:DER_comparison} highlights the predicted value at several buses for our model and the LSTM baseline. 
The constraints on the time-variant matrices acts as a regularizer that allows the model to balance predictions across different buses uniformly, which we observe across all DER levels. In contrast to the LSTM baseline which uses the same optimization criterion, the outputs are inconsistent across buses as the model strives for optimizing the global objective. % 

\begin{table}
\centering
\caption{Comparison of model robustness to sample rate perturbation (MAE $\times10^{-3}$)}
\label{tab:sampling_perturb}
\begin{tabular}{lcccc}
\toprule
& Training & \multicolumn{3}{c}{Evaluation} \\
\cmidrule(lr){2-2} \cmidrule(lr){3-5} 
\makecell{\textbf{Resolution}} & \makecell{\textbf{1 Minute} \\ {\scriptsize(raw 1x)}} & \makecell{\textbf{2 Minute} \\ {\scriptsize(0.5x)}} & \makecell{\textbf{5 Minutes} \\ {\scriptsize(0.2x)}} & \makecell{\textbf{15 Minutes} \\ {\scriptsize(0.06x)}}\\
\midrule
LSTM+SE      & 2.470 & 2.695  & 3.589 & 6.268 \\
\rowcolor{LightBlueOurs} MambaDSSE & 1.661 & 1.672  & 1.714 & 1.784 \\
\bottomrule
\end{tabular}
\end{table}
\begin{figure}
\centering
\includegraphics{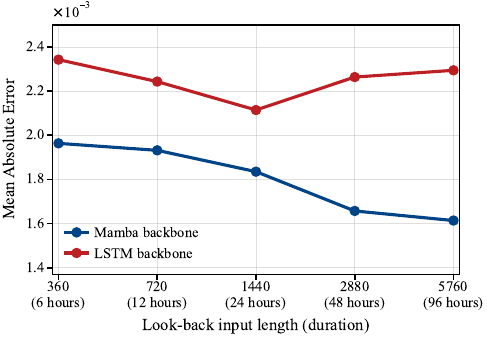}
\caption{Impact of the input sequence length on model performance with a Mamba backbone (blue) and LSTM backbone (red)}
\label{fig:long_rang_dep_ablation}
\end{figure}

We shed light onto two benefits of the selective SSM on the DSSE task. First, we showcase the capacity to capture long range dependencies and its potential utilization for state estimation. Second, we highlight the impact of the learnable step parameter $\Delta_t$ of the input-dependent SSM (Eq.~\eqref{eqn:ssmdelta}) on the models robustness towards data sampling irregularities.
In Table~\ref{tab:sampling_perturb}, we examine the models robustness towards changes in the sample frequency of the data. All the models are trained on 1 minute granularity and tested on irregular sampling rates from 2 minutes (0.5x) upto 15 minute (0.06x). We note that our proposed model mean absolute error remains consistent across different sampling rate perturbations while the LSTM variant doubles, highlighting how the selective SSM gracefully handle irregularly sampled time series and changes in the sampling rate.
In Fig.~\ref{fig:long_rang_dep_ablation} we examine the long range dependency capabilities in Mamba. First we note that the selective SSM backbone improves the models accuracy as we increase the length of the input data. But equally important, utilizing long history structures in the data is contingent on integrating models capable of handling them otherwise, as we observe in the LSTM variant case, longer input lengths can actually harm the performance of the model.

\begin{table}
\centering
\caption{Model runtime performance on different system sizes and input sequence lengths (in milliseconds)}
\label{tab:benchmark}
\setlength{\tabcolsep}{4.6pt}
\begin{tabular}{lcccccccc}
\toprule
& \multicolumn{4}{c}{\textbf{1573 Bus}} & \multicolumn{4}{c}{\textbf{6746 Bus}} \\
\cmidrule(lr){2-5} \cmidrule(l){6-9}
\textbf{Model} & 24h & 48h & 72h & 168h & 24h & 48h & 72h & 168h \\ 
\midrule
LSTM        & 1    & 1.9  & 3.3  & 6.4   & 1.4  & 3.1  & 4.3   & 8.9    \\
LSTM+SE     & 2.37 & 4.43 & 6.77 & 11.3  & 2.57 & 5.19 & 6.86  & 15.31  \\
Mamba Mixer & 1.6  & 2.0  & 3.6  & 6.9   & 1.9  & 3.2  & 4.7   & 9.4    \\
\rowcolor{LightBlueOurs} MambaDSSE & 2.25 & 4.2  & 7.0  & 12.07 & 2.62 & 6.3  & 10.32 & 18.73  \\
\bottomrule
\end{tabular}
\end{table}
Finally, Table~\ref{tab:benchmark} compares the computational performance of the proposed model and baselines on the 1573 and 6746 bus systems with varying input history length 24, 48, 72, and 168 hours. All models are benchmarked on a single A100 GPU, and results reported in milliseconds. We note the impact of integrating a state estimator on the prediction runtime of the models that can be $2\times$ slower than the companion model without an estimator. For example, in the larger 6746 feeder system, the Mamba mixer baseline requires 1.9 ms per prediction and our MambaDSSE model is slowest with 2.52 ms per prediction---mainly due to the iterative filtering process. However, prior works~\cite{Sarkka_2021} have proposed alleviating the sequential nature of the state estimator by exploiting associative properties that provide parallelized filtering implementations to improve the performance and scalability of similar Bayesian estimators for future models.

\vspace{-0.08em}
\section{Conclusion}
\looseness=-1In this paper, we present MambaDSSE, a model-free data-driven method that aims to learn the underlying time-varying behavior of a system to solve distribution system state estimation tasks under high DER penetration levels.
The proposed method combines an input-dependent selective SSM that acts as a matrix generation engine, with probabilistic estimator that operates end-to-end on a lifted measurement space linearized by a learned Koopman observables lifting function.
The model is tested across a diverse set of test systems from the SMART-DS dataset, baseline models, and performance benchmarks.
We demonstrate that the proposed learning framework enables the model to accommodate large system sizes gracefully, and provides robustness towards variations in DER penetration levels, unlike models that employ direct latent-to-state mappings.
We further discuss the contributions of each component to the performance gains, highlighting the role of the Mamba-based backbone in processing long range dependencies, and to the robustness of MambaDSSE to changes in the input sampling rate.

\bibliographystyle{ieeetr}
\bibliography{bib}
\end{document}